%% file: template_ist_ltr.tex
\documentclass[letterpaper,twocolumn,fleqn]{article} 

\usepackage{ist}
\usepackage{tikz}
\usepackage{xspace}
\usepackage{subfig}
\usepackage{url}
\usepackage{hyperref}
 \hypersetup{
     colorlinks=true,
     linkcolor=blue,
     filecolor=blue,
     citecolor = black,
     urlcolor=cyan,
     }

\graphicspath{{figs/}} 

\title{Teaching Color Science to EECS Students Using Interactive Tutorials: Tools and Lessons}

\author{Yuhao Zhu; University of Rochester; Rochester, NY USA}

\date{} 

\begin{document} 

\maketitle 

\thispagestyle{empty} 


\begin{abstract}
Teaching color science to Electrical Engineering and Computer Science (EECS) students is critical to preparing them for advanced topics such as graphics, visualization, imaging, Augmented/Virtual Reality. Color historically receive little attention in EECS curriculum; students find it difficult to grasp basic concepts. This is because today's pedagogical approaches are non-intuitive and lack rigor for teaching color science. We develop a set of interactive tutorials that teach color science to EECS students. Each tutorial is backed up by a mathematically rigorous narrative, but is presented in a form that invites students to participate in developing each concept on their own through visualization tools.
This paper describes the tutorial series we developed and discusses the design decisions we made.
\end{abstract}

\input{macros}
\input{intro}

\input{tutorials}

\input{lessons}
\input{related}
\input{conc}


\small
\bibliographystyle{plain}
\bibliography{refs}


\begin{biography}
Yuhao Zhu received his BS in Computer Science from Beihang University (2010)
and his Ph.D. in Electrical and Computer Engineering from The University of Texas at Austin (2017). He is an Assistant Professor of Computer Science at University of Rochester. His work focuses on applications, algorithms, and systems for visual computing. More about his research can be found at \href{https://horizon-lab.org/}{https://horizon-lab.org/}. Contact him at \href{mailto:yzhu@rochester.edu}{yzhu@rochester.edu}.
\end{biography}

\end{document}

%% file: macros.tex

\newcommand{\website}[1]{{\tt #1}}
\newcommand{\program}[1]{{\tt #1}}
\newcommand{\benchmark}[1]{{\it #1}}
\newcommand{\fixme}[1]{{\textcolor{red}{\textit{#1}}}}

\newcommand*\circled[2]{\tikz[baseline=(char.base)]{
            \node[shape=circle,fill=black,inner sep=1pt] (char) {\textcolor{#1}{{\footnotesize #2}}};}}

\ifx\figurename\undefined \def\figurename{Figure}\fi
\renewcommand{\figurename}{Fig.}
\renewcommand{\paragraph}[1]{\textbf{#1} }
\newcommand{\figline}{{\vspace*{.05in}\hline}}
\newcommand{\funcname}[1]{\textsc{\textbf{\textcolor{blue}{#1}}}}

\newcommand{\Sect}[1]{Section~\ref{#1}}
\newcommand{\Fig}[1]{Figure~\ref{#1}}
\newcommand{\Tbl}[1]{Table.~\ref{#1}}
\newcommand{\Equ}[1]{Equ.~\ref{#1}}
\newcommand{\Apx}[1]{Apdx.~\ref{#1}}
\newcommand{\Alg}[1]{Algorithm~\ref{#1}}

\newcommand{\specialcell}[2][c]{\begin{tabular}[#1]{@{}c@{}}#2\end{tabular}}
\newcommand{\note}[1]{\textcolor{red}{#1}}

\newcommand{\hpm}{\textsc{HPM}\xspace}
\newcommand{\lpm}{\textsc{LPM}\xspace}
\newcommand{\trans}{\textsc{Remeshing}\xspace}

\newcommand{\proj}{\textsc{Mesorasi}\xspace}
\newcommand{\mode}[1]{\underline{\textsc{#1}}\xspace}
\newcommand{\sys}[1]{\underline{\textsc{#1}}}

\newcommand{\RNum}[1]{\uppercase\expandafter{\romannumeral #1\relax}}

\def\cA{{\mathcal{A}}}
\def\cF{{\mathcal{F}}}
\def\cN{{\mathcal{N}}}


%% file: intro.tex
\section{Introduction}

Color science is critical to a wide variety of disciplines and application domains in Computer Science and Engineering (EECS) such as computer graphics, camera imaging, display, information visualization, and Augmented/Virtual Reality (AR/VR). For instance, understanding the physical color formation process is key to understanding how modern digital cameras provide an accurate color perception~\cite{rowlands2020color}, how to develop power-efficient LED displays~\cite{dash2021much, dong2011chameleon, duinkharjav2022color}, and how to develop realistic digital painting applications~\cite{sochorova2021practical}.

As ubiquitous and important as color science is, EECS students find it hard to grasp the essential concepts. Color is usually covered in one lecture in a computer graphics and vision course. Many treat color as a pure sensation; others equate color with the sRGB color cube. Both are unfortunate misunderstandings.

The root causes for a lack of understanding in color science is two-fold. First, while color is a fundamentally 3D concept, they are necessarily visualized in a static, 2D form in classic textbooks~\cite{ware2019information, marschner2018fundamentals, wyszecki1982color}. Students usually stare at one particular 2D projection of a 3D concept without having the ability to interactively examine the different facets of a 3D space. As a result, students lack an intuitive, geometric understand of colors.

Second, while colors are inherently tied to the physical properties of lights and human visual perception, existing tutorials and tools do not present color in a physically, psychophysically, and mathematically rigorous way, leading to confusion when applying color concepts to real-world applications. For instance, while many color tutorials introduce color matching experiments and the resulting Color Matching Functions (CMF), rarely is discussed what exactly the $y$-axis means in a CMF: is it power, or number of photons, or some arbitrary unit? Similarly, the teaching applets by Levoy et al.~\cite{photoapps} (which provides inspiration for our tutorials), while providing certain interactivities to explore color concepts, omit the mathematical derivations.

It must be said, however, that the omission of rigorous physical, psychophysical, and mathematical  derivations in existing color tutorials are well-intentioned. Sometimes the omission is to help students build intuitions without being caught up in details (e.g., in Levoy et al.~\cite{photoapps}). Other tutorials are designed to teach color visualization~\cite{colormatters, harrower2003colorbrewer} rather than building a foundation for research in color science, the latter of which is our goal.

To provide both an intuitive and a physically, psychophysically, and mathematically rigorous treatment of basic color science concepts, we develop a course module on color science, which is offered to both undergraduate and graduate students, the majority of which major in EECS.
The core of our course module a set of interactive tutorials, which use modern visualization packages such as Plotly.js~\cite{plotjs} and Chart.js~\cite{chartjs} to provide intuitive 3D visualization of key color concepts such as chromaticity, color gamut, and imaginary colors that are hard to present in 2D.

Our tutorials, however, are not just a set of visual plots. Instead, each tutorial is designed with a mathematically rigorous narrative to help students derive concepts from first principles step by step. The tutorial content (equations and texts) is dynamically updated based on a student's actions, e.g., the selection of the primary lights. Students see how the choices and assumptions they make lead to different results. Critically, the mathematics (grounded in physics and human visual systems) at each step is explicitly displayed; students can write code to reproduce the derivation of each concept.

Finally, building on the basic color concepts, we also provide tutorials on how to apply color concepts to real-world settings, such as camera color correction and color blindness simulation.

We first present the design guidelines of our tutorials, followed by each one in detail. We discuss the student feedback and lessons we learned, followed by related work. The tutorials are available at \url{https://horizon-lab.org/colorvis/}; the source code will also be released for community development of the tutorials.

%% file: tutorials.tex
\section{Tutorial Design}
\label{sec:tut}

We first discuss the course background and how we prepare students academically for our course module on color science. We then describe the overall design principles of our tutorial series, followed by each tutorial in detail.


\subsection{Course Background and Preparation}
\label{sec:tut:prep}

The course module on color science is a 3-week series in the early part of a larger course that covers various visual computing topics ranging from camera imaging, graphics, display, and AR/VR, all of which rely on basic color science.

Students taking the course are required to have background only in linear algebra and calculus, a reasonable requirement/assumption for a second-year EECS student. No prior exposure to computer vision, image processing, or computer graphics are required. We want to show that the core concepts in color science can be derived from first principles: there is no magic.

We spend two lectures at the beginning of the course introducing the necessary math and physics for learning color science. The first lecture builds on linear algebra and teaches geometric transformation, which is critical to understanding many key color space concepts such as color space transformation and chromaticity. We also teach students what a linear system is (i.e., the additivity and homogeneity), from which we build the ``axiom'' of color science that human color perception is a linear system.

The second lecture builds on basic calculus to  introduce radiometric and photometric properties of light, of which the most important one is the spectral power distribution (SPD) of light. We introduce SPD as the recipe of ``constructing'' a light. We find that many abstract color concepts (e.g., imaginary color) can be intuitively understood in a constructive way, where we ask students: can you construct a light (i.e., specify its SPD) to achieve certain property?

\begin{figure}[t]
  \centering
  \includegraphics[width=\columnwidth]{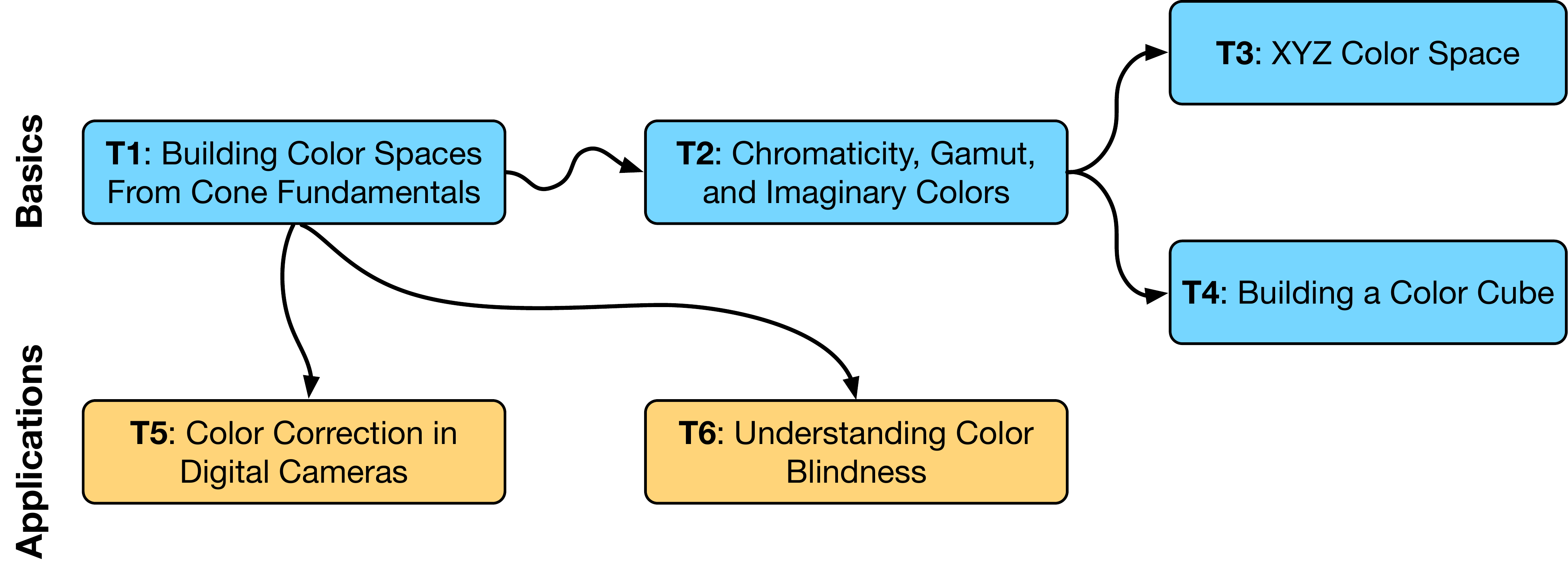}
  \caption{Connections of different tutorials. The first four focus on the basic concepts, and the last two explore color applications.}
  \label{fig:roadmap}
\end{figure}

\begin{figure*}[t]
  \centering
  \includegraphics[width=2.1\columnwidth]{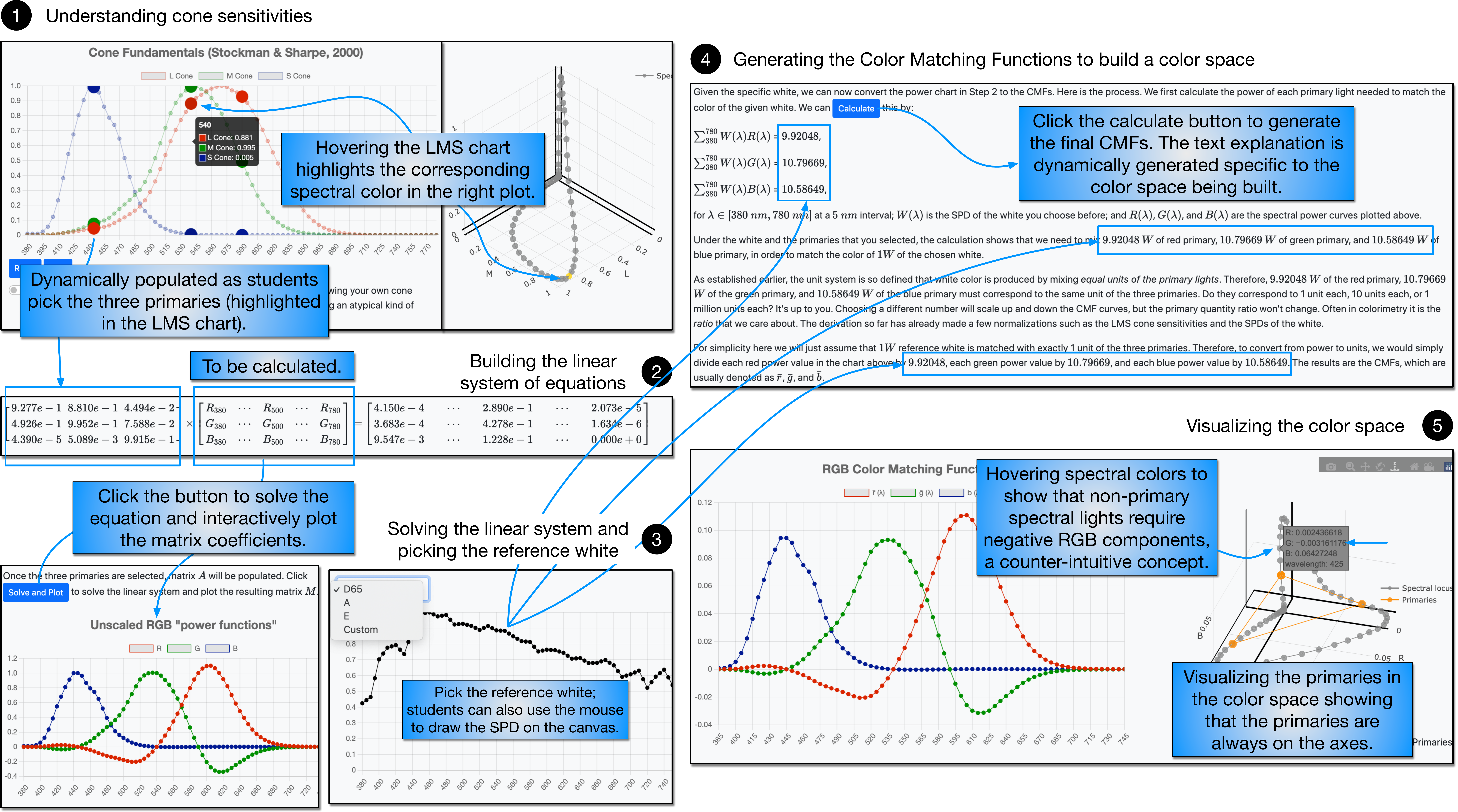}
  \caption{The organization of Tutorial 1, which invites students to build a color space of their own and on their own.}
  \label{fig:t1}
\end{figure*}

\subsection{Overall Design and Principles}
\label{sec:tut:design}

We design the interactive tutorials to complement the classroom lectures on color science. Each tutorial is independent in that one can go through each tutorial without having to have done any previous one, but they are also interconnected in the topics they cover. \Fig{fig:roadmap} shows the relationship of the different tutorials. The first four tutorials cover the basics, and the last two apply the basic concepts to real-world settings. The connections show the dependence in \textit{topics}. For instance, the first tutorial that builds color spaces from cone fundamentals is vital to all other tutorials, since that is where the notion of color is built.

To provide both an intuitive and rigorous understanding of color science, each tutorial is divided into several steps connected by a series of mathematical equations to derive a concept. Each step is associated with interactive plots that not only visualize the concept but provide the flexibility for students to interactively explore different aspects of the concept.

In particular, each tutorial is divided into several steps. Each step starts with a pre-text that explains what they can expect to get out of the step. Students are then asked to perform an action by either interacting with a visual plot (e.g., to pick the primary lights) or pressing a button (e.g., to solve a linear system of equations). As a result of their actions, new data are interactively visualized for them to see the impact of their actions. In the end, we usually have a post-text to explain the results.

While there is a main narrative to drive each tutorial, we also provide freedom for students to explore on their own. Students are invited to experiment uncommon cases to understand the limit of certain concepts. For instance, while the main narrative might use the common D65 light as the reference white, students can draw an arbitrary curve as a white light's SPD. In this way, students see how the choice of white affects a color space and how an extreme choice of white leads to counter-intuitive interpretation of colors.

We usually conclude each tutorial with observations that are otherwise hard to obtain prior to the tutorial. We also intentionally word some of them as teasers for the next tutorial. For instance, once we build a color space we invite students to think about whether a point off the spectral locus correspond to a real color, leading to the discussion of color gamut and imaginary colors in the next tutorial.

We now discuss the first tutorial in great depth to show how the design principles are exercised, followed by other tutorials that are discussed more briefly.

\subsection{T1: Build Color Spaces From Cone Fundamentals}
\label{sec:tut:t1}

\begin{figure*}[t]
  \centering
  \includegraphics[width=2\columnwidth]{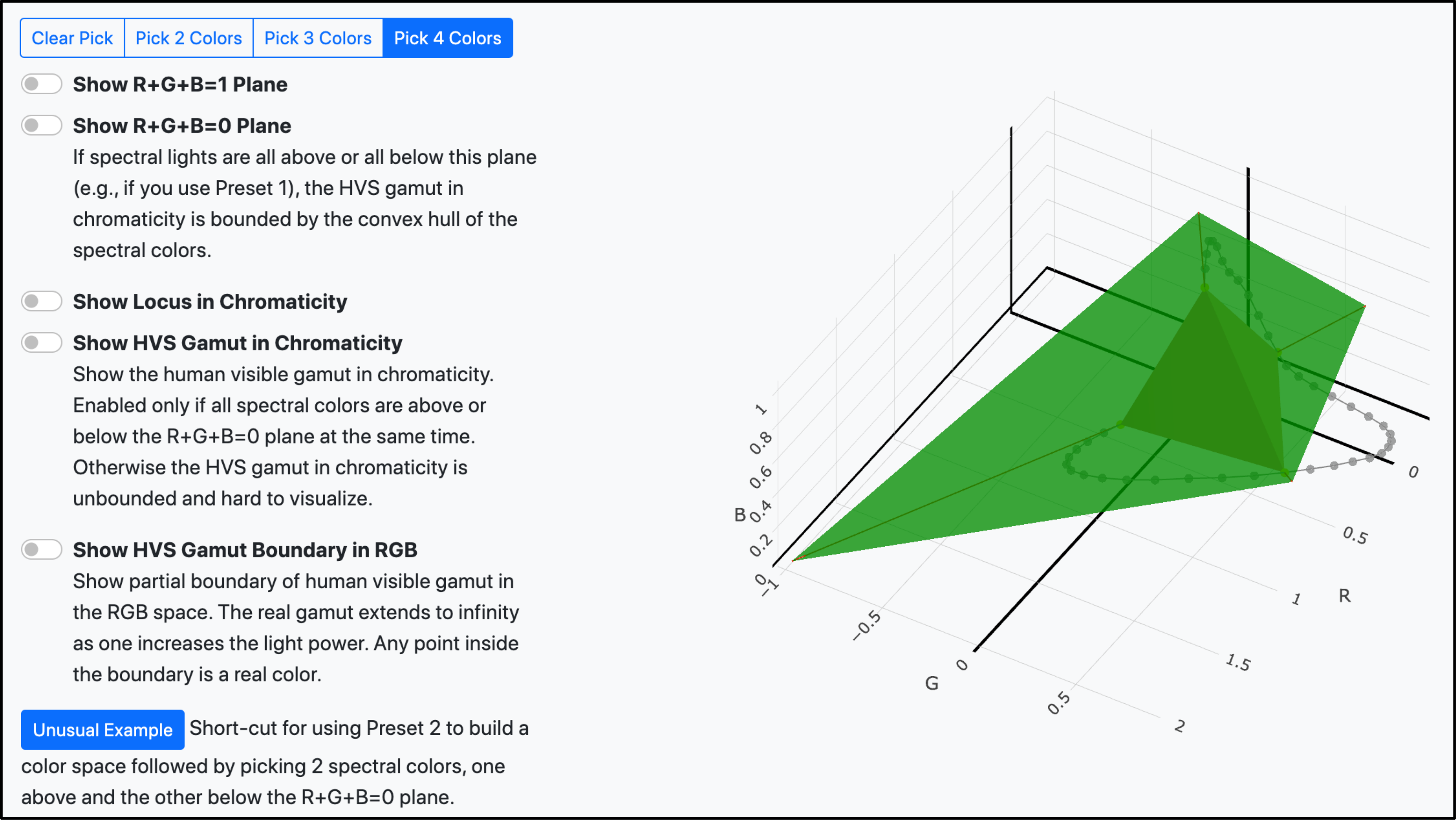}
  \caption{A step in T2, where students are invited to pick arbitrary points (4 here) on the spectral locus as the primaries. We then dynamically show the geometry of all the colors that can be represented by the primaries both in the chromaticity diagram (the green plane) and in the 3D color space (the tetrahedron behind the green plane).}
  \label{fig:t2-1}
\end{figure*}

\begin{figure*}[t]
  \centering
  \includegraphics[width=2.1\columnwidth]{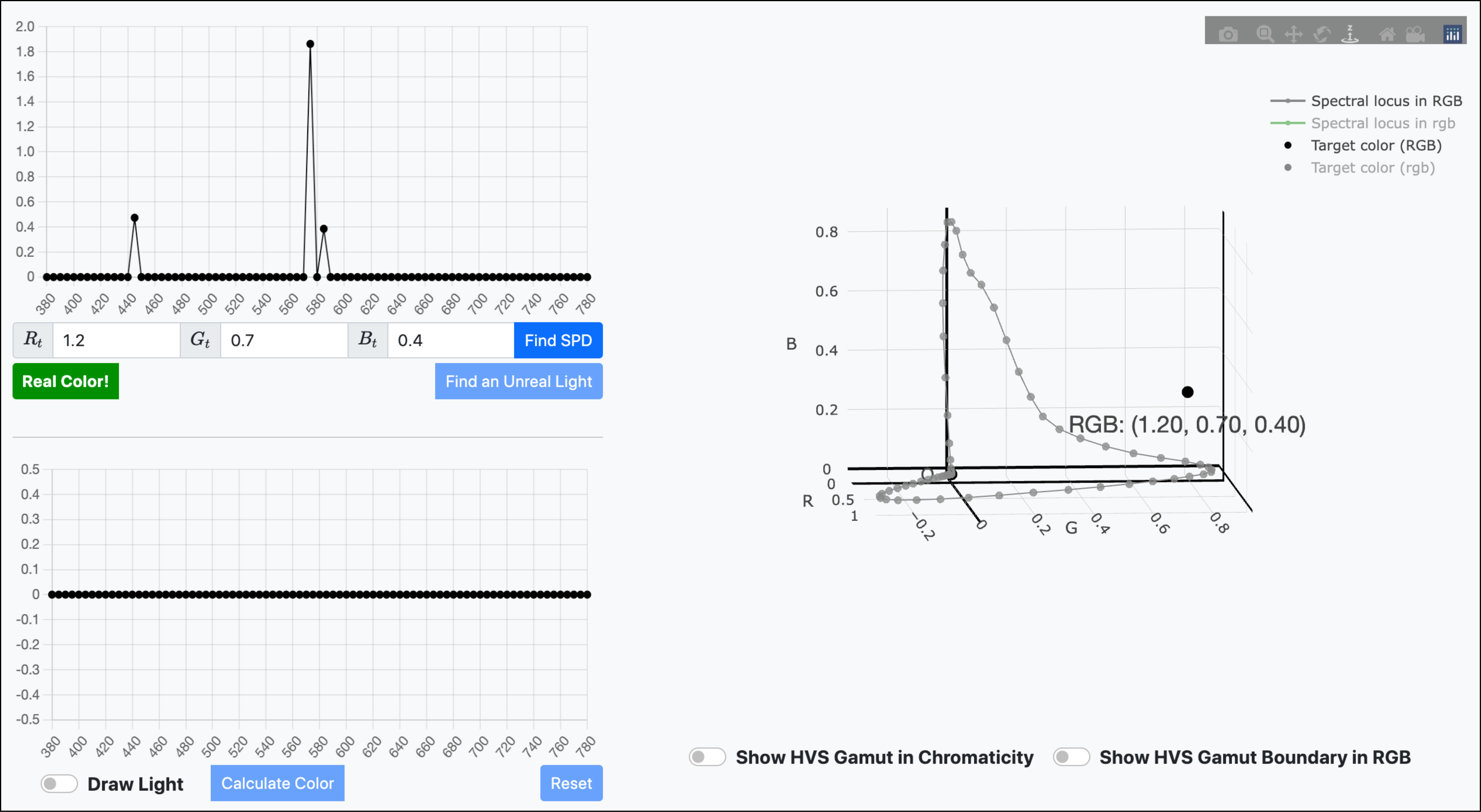}
  \caption{A game in T2, where students are challenged to enter RGB values (the input boxes in the top left panel) that correspond to a real color. They can spin around the spectral locus on the right to build the intuition first. Students can also draw an SPD in the bottom-left panel that is a physically unrealizable light (i.e., has negative components) but that yields a real color.}
  \label{fig:t2-2}
\end{figure*}

\paragraph{Goals.} Students coming into the course have all heard of the vague notion that a color can be represented by three primaries, but they generally do not understand the reason behind it and, in fact, do not know how to state it rigorously. Our first tutorial demystifies this mystery for students, which we find critical to developing a solid understanding in color science.

To that end, the first tutorial\footnote{\url{https://horizon-lab.org/colorvis/cone2cmf.html}} guides the students to \textit{build a color space} of their own and on their own. In particular, they start from the cone fundamentals. While this might seem odd, we believe that it is the right approach, since color matching take places at the cone level, which is the level of abstraction for colorimetry. \Fig{fig:t1} shows the main components in the tutorial.



\paragraph{Main Narrative.} 
We start by showing a 2D plot of the 2-deg cone sensitivity chart, a.k.a., LMS chart~\cite{stockman2000spectral} (left panel of \circled{white}{1}) and provide a rigorous definition: the chart shows the amount of cone responses (y-axis) stimulated from spectral lights (x-axis) \textit{that have the same power} (or any other radiometric measure).


To provide an intuitive understanding of the spectral locus. We show the spectral locus in the 3D LMS space (right panel of \circled{white}{1}) along side the 2D LMS chart. We invite the students to hover the mouse over different spectral lights on the LMS chart, and ask them to notice that a corresponding point on the spectral locus on the right is highlighted as they hover over the 2D chart.

We then ask the students to pick three primaries of their choice by directly clicking three spectral lights on the LMS chart. Once the primaries are picked, a linear system of equation is interactively built and displayed on the webpage (\circled{white}{2}). The linear system shows the key question when building a color space: how much Watts of each primary light do we need to match the color of 1 Watt of each spectral light? The linear system has a form of $A \times M = C$. The matrix $A$ (first matrix in \circled{white}{2}) is dynamically populated as the three primaries are chosen, and the coefficients in the matrix $M$ (second matrix in \circled{white}{2}) are what to be calculated.

Critically, since the equation is dynamically built as the students pick the primaries and they can match the values in the equation with the chart, they get an intuitive understanding of what the equation means: this equation mathematically demonstrates that to match two colors their total cone responses must match.

Students then proceed to click a button to solve the linear system. As a result, matrix $M$ is dynamically plotted (left panel in \circled{white}{3}).
Given the understanding of the linear system of equations (\circled{white}{2}), students know that matrix $M$ must be normalized to convert absolute power to relative units. The ``unit system'' is so defined that equal units of the primaries light should match the color of a reference white, which indicates that the choice of white affects the color space.


To convince students one truly can freely choose the reference white when building a color space, we ask the student to choose a reference white from the CIE Standard Illuminants (right panel in \circled{white}{3}).
The students then click a button to normalize matrix $M$ (\circled{white}{4}), which now represents the Color Matching Function (CMF) of a color space and is plotted in the left panel (\circled{white}{5}).


Through the derivation of the CMFs, students understand the physical meanings behind the CMFs and, thus, appreciate why one must integrate the product of the CMF and the SPD of a light wavelength by wavelength to convert a light to its color. They also understand why a set of CMFs uniquely define a color space: the color of light is a point in the color space described by the CMFs.

\paragraph{Customization.}
First, students can plot custom cone fundamentals (in \circled{white}{1}) to mimic an atypical kind of human being with unique color perception. Second, students can choose primaries that are very close to each other to build a color space (in \circled{white}{1}). They will see that colors in this color space almost lie on a line rather than being spread in the 3D space. Finally, students can also draw an arbitrary SPD and use it as the reference white (in \circled{white}{3}). They will see that when the reference white is, for instance, overwhelming red, any color in that color space will have a strong negative red component.

\paragraph{Post Tutorial Insights.} After constructing a color space, we discuss several interesting observations to provide a deeper understanding.
First, through the equations in \circled{white}{2} students get to have a mathematical explanation as to why human vision is trichromatic: the system of equations would be under- or over-determined if less or more than three primaries were used.

Second, by hovering over all the spectral lights in the color space they just built (right panel in \circled{white}{5}), students see that all spectral lights (other than the three primaries) have at least a negative component. Students are encourage to change the primaries to see if this conclusion holds. This is a teaser for the students to starting thinking about whether there exists a color space that can produce the colors of all the spectral lights at the same time, a natural lead-in to color gamut in the next tutorial.

\subsection{T2: Gamut and Imaginary Colors}
\label{sec:tut:t2}

\paragraph{Goals.} Following the first tutorial where students learn to build a color space on their own, the second tutorial\footnote{\url{https://horizon-lab.org/colorvis/chromaticity.html}} teaches in color science: gamut of Human Visual System (HVS), which naturally gives rise to the notion of imaginary colors.

 

\paragraph{Main Narrative.} We start with a question: what are all the colors that the HVS can see? To answer this question, one could construct every single real light and find its color in a color space. While we cannot really do that in practice, constructing a few lights that way helps the students build the intuition.

To that end, we build an interactive plot that shows the spectral locus and ask student to pick up to four arbitrary spectral lights. See the students view in \Fig{fig:t2-1}. Dynamically, we show the colors that can be constructed from mixing those spectral lights. For instance, when picking four lights, we visualize a tetrahedron; any point within that tetrahedron represents a color that is produced from mixing the four chosen spectral lights. By trying different combinations, students get an intuitive sense of what sort of colors can be made from real lights.

We then introduce the notion of HVS gamut---all the colors that real lights generates.
From the interaction above, students sense that the HVS gamut occupies only a subset of an RGB space. A natural question is: what about an arbitrary point in an RGB space; does it correspond to a real color; if so, what light generates that color?

Our tutorial builds a mini game to allow students to explore the answer of this question. \Fig{fig:t2-2} shows the students' view. Students are challenged to pick a point in the RGB space that represents an imaginary color (by entering the three values in the top-left panel). Students can build their intuition by first spinning around the spectral locus on the right to see what colors/points might lie inside the HVS gamut.

Under the surface, the tutorial will either visualize a SPD that could have led to that particular RGB color---if the color is a real color (the case shown in \Fig{fig:t2-2})---or, otherwise, plot the SPD of a physically-unrealizable light that mathematically gives the target RGB color but with negative SPD values.

\paragraph{Customization.} Students can build a color space that challenges the conventional wisdom that the HVS gamut is the area under the convex hull of the spectral locus in chromaticity. In particular, if the primaries are chosen in such a way that when the spectral colors are separated by the $R+G+B=0$ plane, the HVS gamut extends to infinity in the chromaticity plot, which the students can visually see.

\paragraph{Post Tutorial Insights.} Students see that the LMS spectral locus resides completely in the positive octane of the LMS space. This indicates that any point on the axes is out of the HVS gamut, indicating that the LMS primaries are imaginary colors.

Interestingly but perhaps unsurprisingly, it is possible that a real light and a physically-unrealizable light are metamers (i.e., corresponding to the same color). To verify that, students draw an arbitrary SPD with negative components (usually with many positive components and a few negative components). This is done in the bottom-left panel in \Fig{fig:t2-2} (visualization results not shown here). We then show whether it has a metamer that is a real light.

\subsection{T3: XYZ Color Space}
\label{sec:tut:t3}

\paragraph{Goals.} This tutorial\footnote{\url{https://horizon-lab.org/colorvis/xyz.html}} introduces the CIE XYZ color space, critical for all subsequent discussions in color science, as the XYZ space is the ``common language'' for color space transformation.

From our lectures, students generally understand that CIE XYZ has the nice feature that all real colors have positive primary contributions, and is simply a linear transformation from an RGB color space. What the students find hard to grasp is the notion that there are many such transformations but the CIE 1931 standard chose a particular one---what is the rationale?

\paragraph{Main Narrative.} Starting the CIE 1931 RGB color space, we first visually show the transformation from RGB to XYZ without discussing the mathematical derivation. Students see that the HVS gamut resides completely within the positive octant in the XYZ space (an important requirement of the XYZ space), and there exists a linear transformation from RGB to XYZ.

Next, we state the problem at hand: finding the nine matrix coefficients in $T$ to transform any given $[R, G, B]$ triplet to a corresponding $[X, Y, Z]$ triplet. We then, in three steps, progressively introduce the constraints that CIE 1931 imposes~\cite{fairman1997cie, brill1998cie}; each constraint allows us to solve for a subset of matrix $T$'s coefficients.

\subsection{T4: Building a Color Cube}
\label{sec:tut:t4}

\paragraph{Goals.} Students coming into the course, even without formal exposure color science, know that a common way to specify color (e.g., in a graphing application) is by specifying the R, G, and B value of color, each of which is an integer within the range of 0 and 255. This idea conflicts with what they have learned so far: for all they know a color might even have negative primary components!
This tutorial\footnote{\url{https://horizon-lab.org/colorvis/colorcube.html}} connects how a color scientist describes colors to how an application developer describes color---by guiding them to build a color cube for a particular color space.

\begin{figure}[t]
  \centering
  \includegraphics[width=\columnwidth]{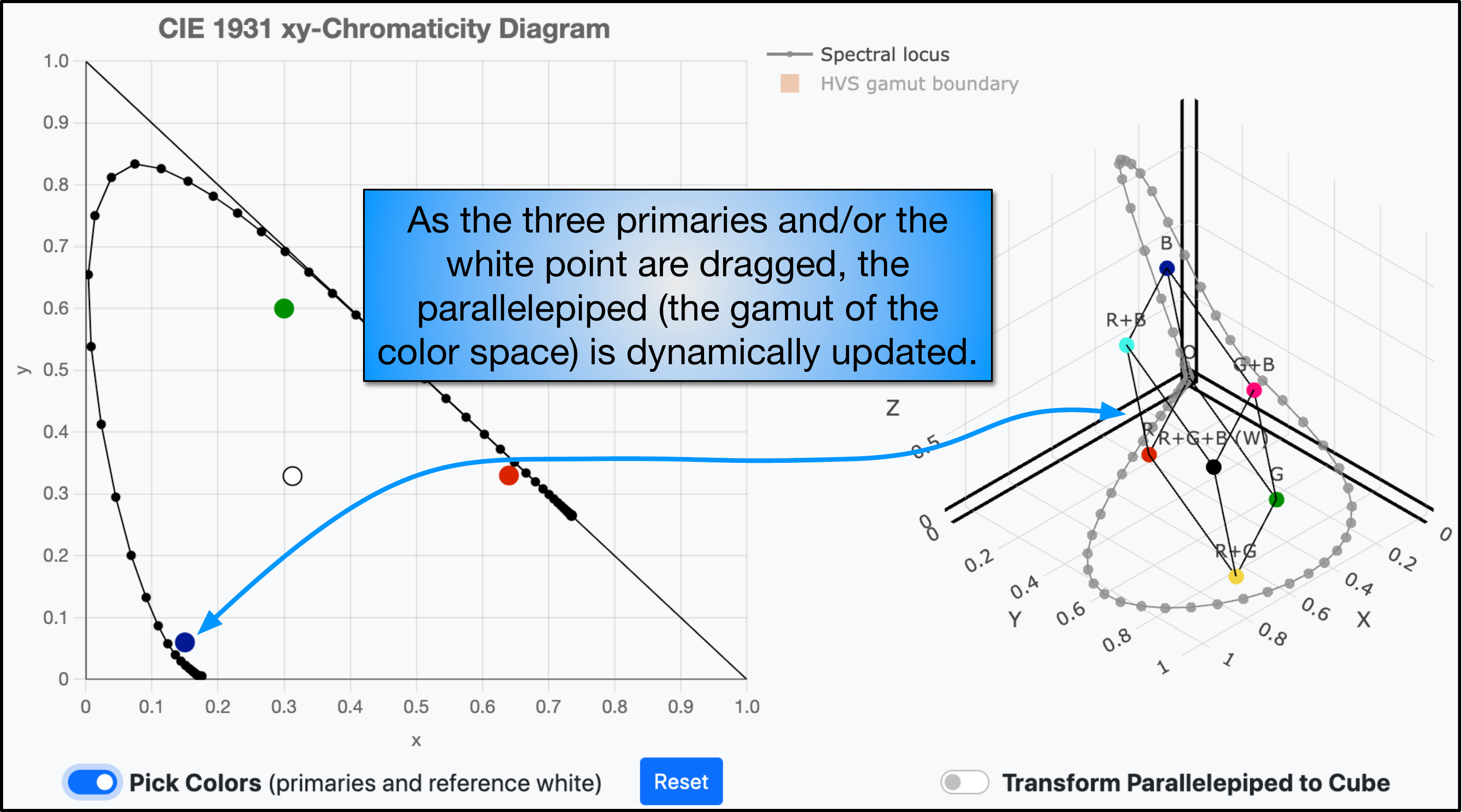}
  \caption{A step in T4, where students learn to build a color cube from a color space. As the primaries/white point are dragged on the left, the color gamut (the parallelepiped on the right) is dynamically updated.}
  \label{fig:t4}
\end{figure}

\paragraph{Main Narrative.} We divide it into three three steps to build a color cube. First, students get to see that the gamut of an RGB color space (colors that are expressed as non-negative combination of the primaries) is a parallelepiped. They can drag the three primaries and the reference white to change the RGB color space. Dynamically, the corresponding parallelepiped is dynamically changed accordingly, as \Fig{fig:t4} shows. Second, students transform a parallelepiped to a cube (a linear transformation). Finally, we introduce the quantization process, during which we discuss gamma.
The mathematical derivation of each step is laid out with parameters dynamically updated depending a student's choice of the RGB color space.

\paragraph{Customization.} Our visualization tool allows students to build counter-intuitive RGB cubes. For instance, students can use imaginary colors as the primaries. In this case, they will see that the resulting color cube will go outside the HVS gamut, indicating that some colors expressed in the color cube are imaginary colors. Another experiment students try is to set the reference white to be outside the RGB triangle in the chromaticity plot (left panel in \Fig{fig:t4}), which indicates that the reference white is not positively mixed from the primaries. This is mathematically sound, but the resulting color cube will be again outside the HVS gamut.

\subsection{T5: Color Correction in Digital Cameras}
\label{sec:tut:t5}

\begin{figure*}[t]
  \centering
  \includegraphics[width=2.1\columnwidth]{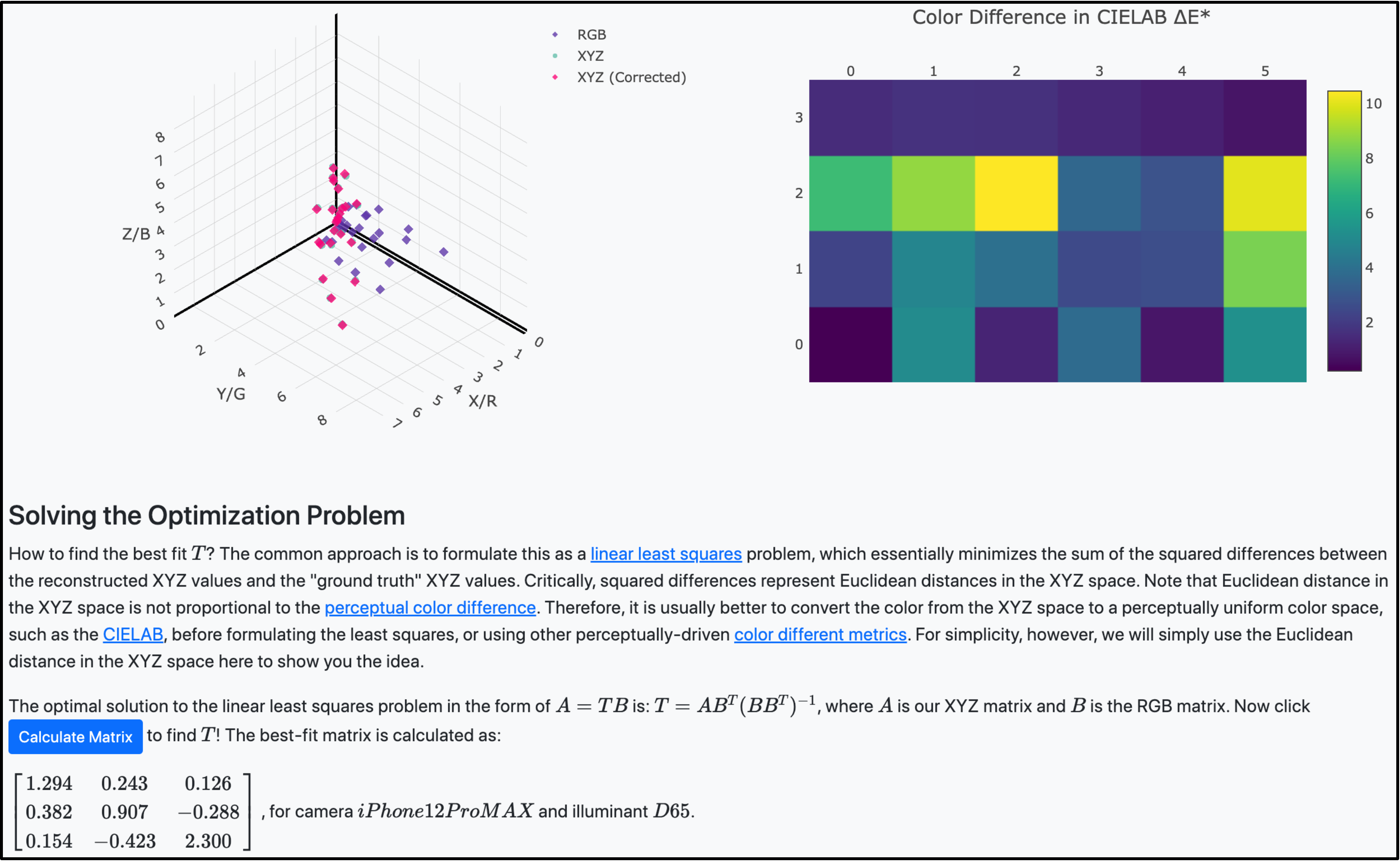}
  \caption{A step in T5, where we show the original RGB values and the corresponding XYZ values of the calibration patches. We discuss that color correction is an optimization problem that estimates the best-fit linear transformation (see the matrix). We also show the CIELAB $\Delta E^*$ metric, from which students understand that color correction is not perfect. Critically, all the visualization and information here will be dynamically updated as the students choose a different camera, a different illuminant, or a different set of calibration patches.}
  \label{fig:t5-1}
\end{figure*}

\paragraph{Goals.} Now that students have a solid understanding of the basic concepts in color science, the next two tutorials, T5 and T6, apply the basic concepts they have learned to real-world applications. T5\footnote{\url{https://horizon-lab.org/colorvis/camcolor.html}} discusses how cameras (attempt to) reproduce colors as close to our human color perception.

\paragraph{Main Narrative.} Our strategy is to make an analogy between a camera and our HVS. Normally, each camera has three color filters, each of which has a spectral sensitivity to lights, just like the cone fundamentals of our eyes. We visualize the measured camera sensitivities from 48 recent cameras (including smartphones, DSLRs, and industrial cameras) ~\cite{jiang2013space, tominaga2021measurement} in a plot and overlap the LMS cone sensitivities. Students immediately see that the shapes of the camera sensitivity functions more or less resemble, but do not match exactly, those of the cone fundamentals.

We then state our goal: finding a linear transformation from the camera sensitivities to the cone sensitivities.
This problem is translated to solve for the nine coefficients in the transformation matrix. \Fig{fig:t5-1} shows the students' view when trying to understand this problem.
Using the ColorChecker as the calibration patches~\cite{babelcc}, we show the optimization formulation, which is a linear least squared problem.
After clicking the ``Calculate Matrix'' button, the best-fit transformation matrix is displayed.

After finding the matrix, we visualize in one plot the original, corrected, and the ground truth colors of the calibration patches. On the right we show the correction accuracy using the CIELAB $\Delta E^*$ metric. Students see that the correct values align well with the ground truth values. They, however, do not exactly overlap, because the estimated transformation matrix is not a perfect fit.

\begin{figure}[t]
  \centering
  \includegraphics[width=\columnwidth]{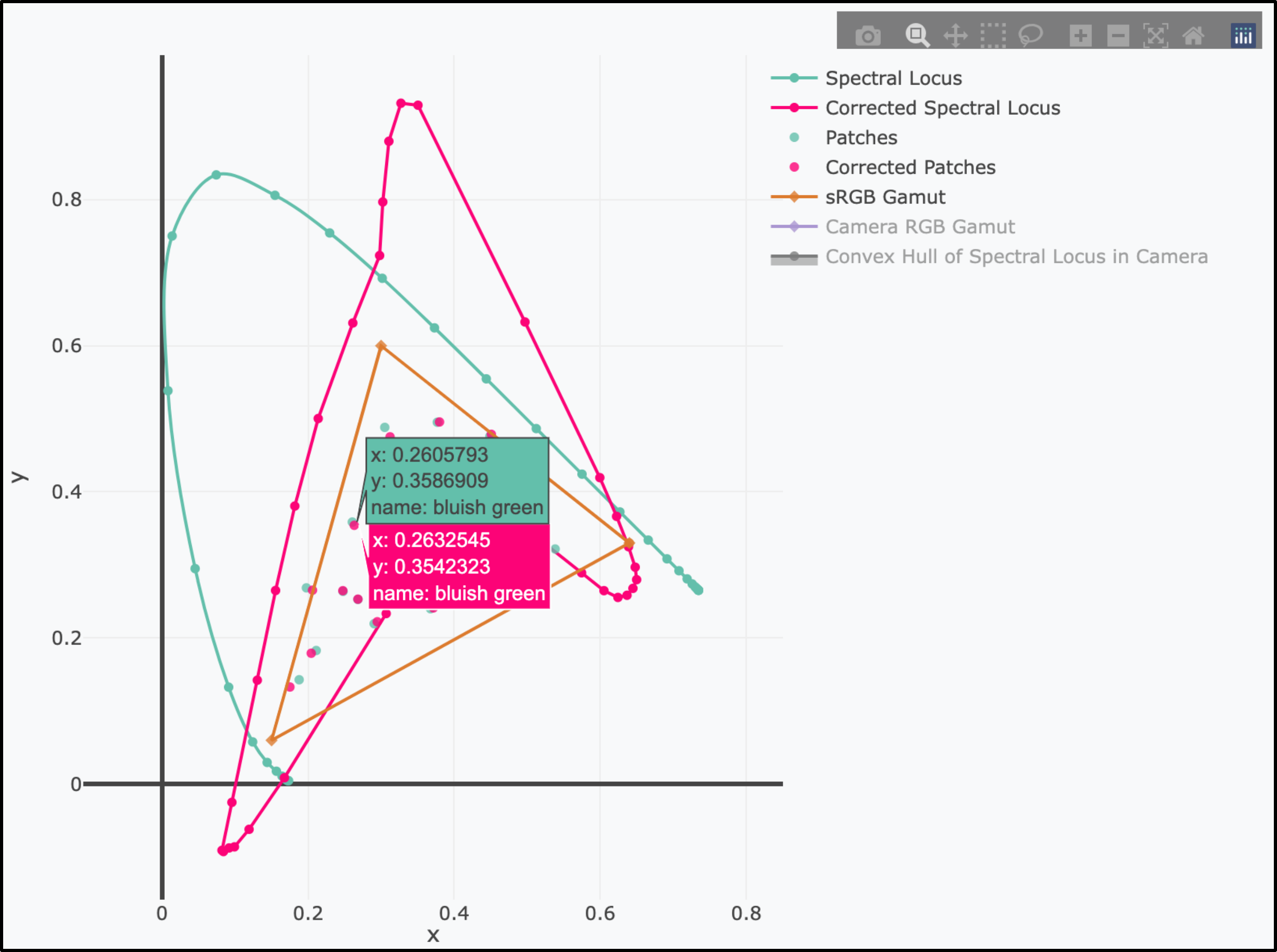}
  \caption{Part of T5, showing the spectral locus and the calibration patches after correction.}
  \label{fig:t5-2}
\end{figure}

\begin{figure*}[t]
  \centering
  \includegraphics[width=2.1\columnwidth]{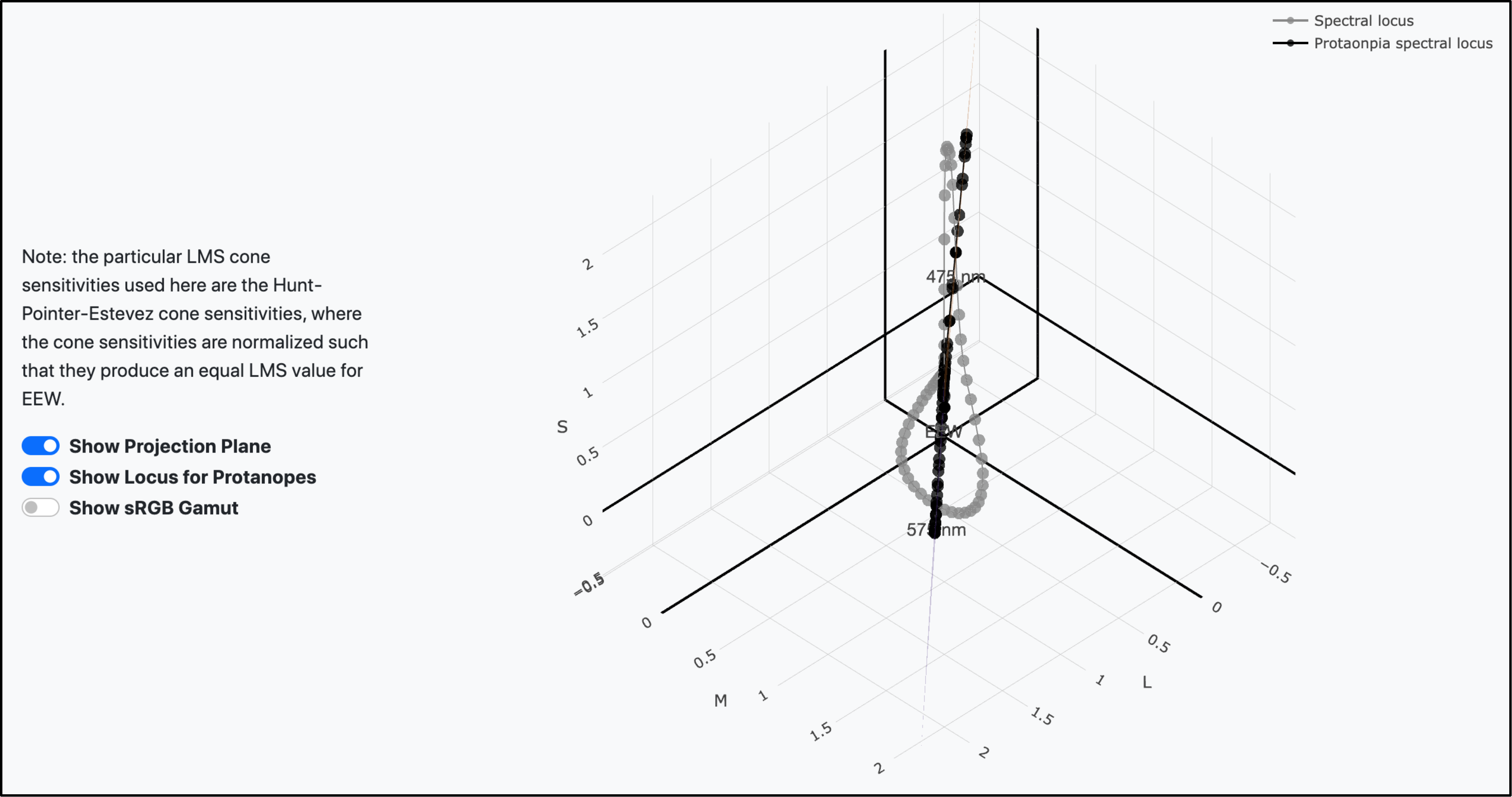}
  \caption{A step in T6, where we visually show that a dichromat's color vision can be seen as projecting colors that a trichromat can see to a plane. The visualization here shows the original spectral locus and the projected locus.}
  \label{fig:t6}
\end{figure*}

\paragraph{Customization.} Students can draw camera sensitivity functions that mimic the LMS cone fundamentals or the XYZ CMFs. They see that the color reproduction error are relatively small. They can also choose a different camera or illuminant and see that color correction is necessarily camera/illuminant-dependent and is harder on some cameras, where a linear transformation is harder to find. Critically, the view in \Fig{fig:t5-1} will be dynamically updated as the students choose different options.

\paragraph{Post Tutorial Insights.} Students learn that the colors of spectral lights have large errors, because spectral lights are not used as the calibration targets. Students' view is shown in \Fig{fig:t5-2}. This is evident when they see that the actual spectral locus and the corrected spectral locus do not overlap at all. When the mouse hovers upon a uncorrected color patch (cyan in \Fig{fig:t5-2}), the corresponding corrected color will be highlighted as well (pink here), showing the correction quality.

Students can also see that the corrected spectral locus as captured by the camera (the pink curve in \Fig{fig:t5-2}) is no longer convex. Therefore, it is no longer true to say that all real colors lie inside the locus. It is in fact the convex hull of the corrected locus that contains all the real colors.

Last but not least, students can see that the ColorChecker colors under D65 lie mostly within the sRGB color gamut (orange triangle in \Fig{fig:t5-2}), which is intentional: sRGB is the most commonly used color space, and one would want to be able to reproduce its gamut well.

\subsection{T6: Understanding Color Blindness}
\label{sec:tut:t6}

\paragraph{Goals.} This tutorial\footnote{\url{https://horizon-lab.org/colorvis/colorblind.html}} discusses color deficiency. We use the theory described by Brettel et al.~\cite{brettel1997computerized} to show how to simulate the color perception of a dichromat (using protanopia as an example).

\paragraph{Main Narrative.} Many online tools exist that simulate color blindness~\cite{coblis, colororacle}, but the underlying color scientific principles are not very well discussed in those tools (with the exception of DaltonLens~\cite{daltonlens}), which our tutorial attempts to deliver.

We first build the intuition that the color perception of a dichromat, rather than being 3D, is 2D since one cone type is missing. To that end, we show that geometrically one can think of a dichromatic vision as projecting a point $[L, M, S]$, a color that a trichromat can see, onto a plane $\mathbf{P}$ (given by Brettel et al.~\cite{brettel1997computerized}); the projected point $[L', M', S']$ represents the actual color that a dichromat would see. \Fig{fig:t6} shows the students view, where we project the spectral locus to the plane $\mathbf{P}$. In the narrative (not shown here), we mathematically derive the plane\footnote{Actually Brettel et al. uses two planes, a detail that we discuss in our tutorial but is omitted here.}.

Given the assumption $M'=M$ and $S'=S$ in a protanope (since only the L cones are missing), students see that this projecting is essentially an orthographic projection of the point $[L, M, S]$ toward the $MS$-plane and intersecting the projection with the plane $\mathbf{P}$. Equivalently, students also see that this is an oblique projection of the point $[L, M, S]$ onto $\mathbf{P}$.

Given the theoretical understanding, students upload any image (presumably encoded in sRGB color space) and go through the protanopia simulation to see how that image is seen by a protanope. We also visualize, in the XYZ space, the pixel colors seen by both a trichromat and a protanope. Students see that the pixel colors perceived by a protanope all lie on a single plane.

\paragraph{Post Tutorial Insights.} Students see that the white color in an sRGB-encoded image appears pink-ish. This is because sRGB white, after protanopic projection, is outside the sRGB gamut. Students can visually convince themselves of this using the visualization tool in the tutorial, where the sRGB white when projected to the plane $\mathbf{P}$ is outside the HVS gamut. Our simulation simply clamps the values to the [0, 255] range, essentially performing a gamut mapping using the Absolute Colorimetric rendering intent~\cite{morovivc2008color}.

%% file: lessons.tex
\section{Student Feedback and Lessons Learned}
\label{sec:lessons}

\paragraph{Feedback.} We conducted an informal survey from the students about the tutorials, and the feedback was positive. Among about 30 students who took the course, over two-thirds of them found that the tutorials deepened understanding.

We conducted a pre-tutorial and post-tutorial survey on T5 (camera color correction), both after the classroom lecture on color correction. Both surveys asked students to briefly describe the key principle of color correction. Fewer than 10 students provided a clear answer before taking the tutorial and almost all students provided the correct answer after going through the tutorial.

Two additional evidences suggest the positive impact of the tutorials. Some students took on final course projects to reproduce the math behind several tutorials. They cite in their final report that the tutorials help their implementations. In addition, some students find the tutorials useful to the point that they help design additional tutorials. For instance, the color blindness tutorial is partially designed by a former student used the tutorials before.

\paragraph{Lessons.}
First, students appreciate being able to interact with 3D plots to understand abstract color concepts such as imaginary color and HVS gamut, which is made easy when students see where a color sits along with the spectral locus.

Second, it is vital to show students physically-rigorous math, which should be interactively derived as students interact with the tutorial. For instance, when deriving the CMFs the coefficients in the system of linear equations are dynamically populated based on students' choice of primaries. The dynamic update provides an important cue for students to understand the matrix.

Third, providing visualization tools that allows students to deviate away from the main narrative of a tutorial deepens their understanding. For instance, by allowing students to draw custom primaries in T2, students get to see the spectral locus is not necessarily convex in any given RGB color space.



%% file: related.tex
\section{Related Work}
\label{sec:rw}

Many excellent online interactive tools exist for exploring color concepts. Most of them, however, focus on \textit{color visualization} (e.g., how to use color for better visualization), rather than developing concepts of \textit{color science} from a physical and psychophysical perspectives. Notable examples include ColorBrewer~\cite{harrower2003colorbrewer, colorbrewer2} for picking colors for cartography, David Johnstone's Lch and Lab colour and gradient picker~\cite{gradpicker}, CCCTool~\cite{ccctool} and ColorMoves~\cite{colormoves} for constructing colormaps, Moreland~\cite{moreland2016we, colormapadvice} and Zhou and Hanson \cite{zhou2015survey} provide comprehensive discussions and surveys of colormapping.

Bartosz Ciechanowski~\cite{csvis} and Jamie Wong~\cite{hex2eye}, independently, build many celebrated interactive visualizations for exploring colors. Apart from covering (many) more concepts in color science, our tutorials provide rigorous derivations.

%% file: conc.tex
\section{Conclusion}
\label{sec:conc}

Having a solid understanding of color science basics is critical to developing further insights in many visual computing domains. We describe how visualization tools combined with interactive tutorials allow students to grasp abstract concepts and deepen the understanding in color science. Central to our tutorial series is the rigorous derivations of color concepts from cone fundamentals and physical attributes of light. In this way, students get the sense that there is no magic; everything can be derived from first principles. Students build confidence through the color literacy they obtain from these derivations.

Our tutorials are also interactive. While many visualization plots can easily be found in any color science textbooks, students have historically learned from printed and static versions of these plots. Interaction allows students to explore different aspects of these ``famous'' diagrams and better understand the concepts. Finally, we let students customize their learning experience (e.g., draw custom primary lights), drawing insights on their own. The tutorial series is continuously being developed and we hope to build community efforts around it.